\documentclass[esprc2,12pt]{article}
\usepackage{graphicx}
\usepackage[top=3cm, bottom=2cm, left=3cm, right=3cm]{geometry}

\title{Cosmic ray antiprotons and the Single Source model} 
\author{$^{*}$A.D. Erlykin $^{1,2}$ and A.W. Wolfendale $^{2}$\\
$(1)$ P N Lebedev Physical Institute, Moscow, Russia.\\
$(2)$ Physics Department, Durham University,\\ Durham, DH1 3LE, UK}

\begin{document}
\maketitle

\footnote{$^{*}$Corresponding author: tel +74991358737 \\ 
 E-mail address: erlykin@sci.lebedev.ru}

\begin{abstract}
In view of the fact that the AMS-02 instrument has recently been used to make 
preliminary observations of the ratio of the antiprotons ($\bar{P}$) to protons (P)
 in the primary cosmic radiation we have returned to our idea of signatures of 
a local recent supernova. We find that at the present level of accuracy there 
is no inconsistency between our predicitions for the $\bar{P}/P$ ratio to some 
hundreds of GeV using the preliminary observations.
\end{abstract}

Keywords: cosmic rays, antiprotons, protons
\section{Introduction}
One of the most striking puzzles in cosmology is the baryon asymmetry of the Universe,
i.e. the absence of antimatter in the visible part of it. Many experiments have been 
carried out to search for antimatter in primary cosmic rays (CR). So far all of them 
have given negative results. All the observed antimatter particles:  positrons and 
antiprotons have been found to have a secondary origin, i.e. they are produced in the 
interactions of primary CR with the particles of the Interstellar Medium (ISM). 

A most exciting discovery 
was made recently by the PAMELA collaboration, who showed that the energy spectrum and
positron to electron ratio have an unusual energy dependence and cannot be explained
by the positrons being entirely secondary particles\cite{Adriani}. This experimental 
result has 
been confirmed later by the Fermi-LAT \cite{Fermi} and, with high precision, by the 
AMS-02 \cite{Aguilar} experiments. Many models have been proposed to explain the 
observations and among them our model of the Single Source \cite{EW1}.

Initially the Single Source model was proposed for the explanation of the evident
sharpness of the knee in the primary CR energy spectrum at an energy of about 3-4 PeV 
\cite{EW2}. It was based on the obvious non-uniformity of the ISM and of the 
distribution of stars in space. It assumes that the knee is caused by the contribution 
of a nearby and recent supernova remnant (SNR). We think that the most likely candidate
 for such a SNR is Vela SNR \cite{EW3}.

The excess of positrons is observed at an energy by 4 orders of magnitude less than 
that of the knee, so that the source of the positron excess should be different from 
that responsible for the knee. We suppose that the most likely candidate for such a 
positron source in the sub-TeV energy region is the Geminga SNR. We made the case for 
the positrons being derived from radioactive nuclei generated in the SN explosion 
\cite{EW1}.

Recently the AMS-02 collaboration have presented the preliminary results of the search 
for another antimatter particle - the antiproton. They show that in about the same 
sub-TeV energy region: from 20 GeV to 450 GeV, the $\bar{P}/P$ ratio stays constant 
\cite{AMS1}. This behaviour cannot be explained by the secondary production of 
antiprotons from ordinary CR collisions.

In what follows we examine the situation with CR antiprotons on the basis of the 
Single Source model, i.e. the possibility of explaining the unusual behaviour of the
\^{P}$/$P ratio by the contribution of the recent nearby SNR. 
\section{The antiproton - proton ratio}
\subsection{General remarks}
As with the situation for positrons the flux measured at Earth comprises a background 
component, due to CR interactions in the ISM and a (possible) component due to a 
discrete source. The latter component is very clear in the positron case but less so 
for antiprotons, as yet.
The background is a crucial part of the analysis and this will be considered first. 
\subsection{The antiproton background spectrum}
Many factors contribute to causing significant uncertainty in the background (see 
\cite{EW1,Gies} for details). We ourselves have made an independent estimate, being
careful to include the contribution from helium nuclei (He) in the CR and in the 
target interstellar medium (ISM). 

Our analysis is based on the model calculations of antiproton production in PP, PA and 
AA - interactions made using EPOS-LHC and QGSJET-II-04 codes \cite{Kach}. The advantage
 of this work is that it uses the most advanced Monte Carlo generators tuned to 
numerous accelerator data including those from the Large Hadron Collider. The sample of
projectile and target nuclei covers all major CR nuclei: P, He, CNO, Mg-Si and Fe for
projectiles and P, He for targets. The energy range covered by these calculations is 
from 1 to 10$^4$GeV/nucleon, i.e. it is adequate for the analysis of the available
experimental data. The results are given in numerical tables which are useful for
the precise calculations.

The numerical results of this work are given for so called Z-factors. They are 
defined as spectrally averaged energy fractions transferred to antiprotons assuming 
that the spectra of CR species in the relevant energy range can be approximated by a 
power law: $I_i(E) = KE^{-\alpha}$. Values of the spectral index $\alpha$ vary from 2.0
 to 3.0. The Z-factors are expressed via the inclusive cross-section for the 
production of antiprotons $d\sigma /dz$, $z = E_{\bar{P}}/P$, as 
$Z = \int_{0}^{1} dz z^{\alpha-1} \frac{d\sigma}{dz}$. The contribution $q^{ij}$ of the
 particular inclusive reaction $i+j \rightarrow \bar{P}+X$ to the antiproton flux can 
be calculated as $q^{ij} = n_jI_iZ^{ij}$. Here, $n_j$ is the number column density of 
the target nuclei. We adopted an ISM consisting of protons (70\%) and helium nuclei 
(25\%); heavier nuclei contribute no more than a few percent of the total number.

In order to convert inclusive cross-sections to inclusive spectra of produced 
antiprotons, the Z-factors were multiplied by the column density of the interstellar 
gas passed by the projectile nucleus (P or He) during its life in the Galaxy. It is 
taken as $cT\rho$, where $c$ is the speed of light, $T$ is the life time of the 
projectile particle in the Galaxy and $\rho = 0.5cm^{-3}$ is the density of the ISM.
The life time $T$ depends on the rigidity $R$ of the particle as 
$T = \tau_0 \cdot R^{-0.5}$, where for $R = 1 GV$  $\tau_0 = 4\cdot 10^7$year. After 
the antiprotons are produced they begin to diffuse and finally annihilate or escape 
from the Galaxy. We assumed that their life time in the Galaxy against the escape is 
the same as for protons. The annihilation cross-section for antiprotons is taken
from approximations used in the CORSIKA6500 code. 

The results of our calculations are shown in Figure 1. They are compared with the 
$\bar{P}/P$ ratio measured in the PAMELA and AMS-02 experiments \cite{Adriani,AMS1}.
\begin{figure}[htb]
\begin{center}
\includegraphics[height=15cm,width=9cm,angle=-90]{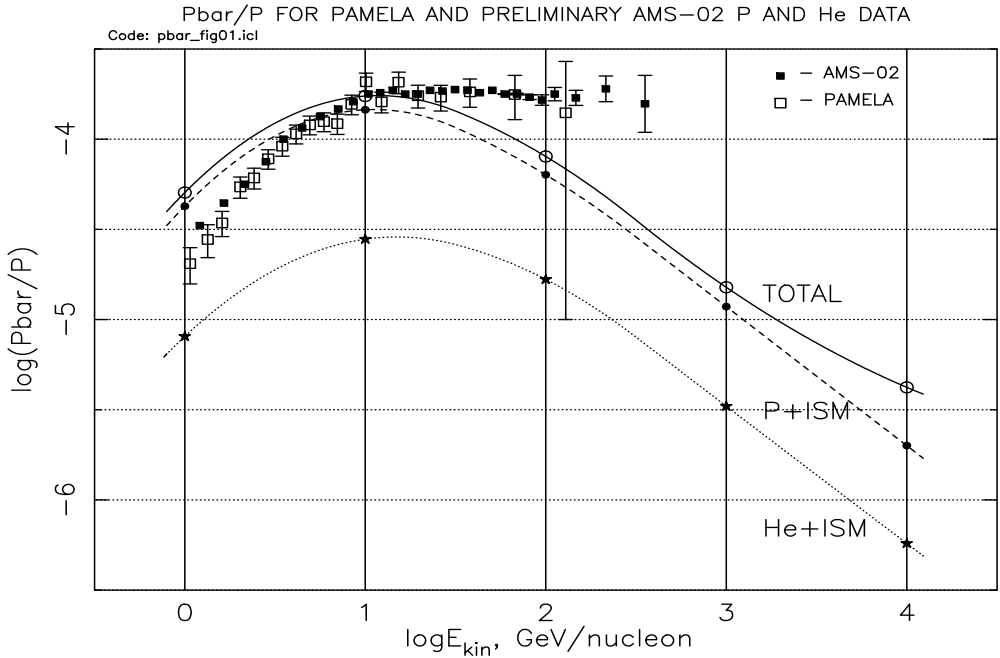}
\end{center}
\caption{\footnotesize The antiproton to proton ratio measured by the PAMELA 
\cite{Adriani}
(open squares) and AMS-02 \cite{AMS1} (full squares) experiments compared with the 
calculations. Contributions from proton and helium nuclei interactions with the ISM 
(P+He) as well as their sum are shown as dashed, dotted and full lines; denoted as
P+ISM, He+ISM and TOTAL respectively.}   
\label{fig:fig1}
\end{figure}   
Contributions from P+ISM and He+ISM reactions are given separately and their sum is 
plotted by the full line. It is seen that the contribution of He-induced reactions does
 not exceed $\sim$30\% of the total. No normalisation is applied in the comparison and 
the excess of the experimental $\bar{P}/P$ ratio over the calculated background is 
clearly seen at energies above 20 GeV/nucleon. 

The experimental uncertainties are shown by the 
vertical error bars. The uncertainty of the calculations is difficult to estimate 
because, as it has been shown above, there are many input parameters with their own 
uncertainties. Some of the uncertainties are reduced in the ratio of $\bar{P}$ to $P$,
 in comparison with the actual separate particle intensities, but not all. 
It can be noted that the resulting discrepancy at 1 GeV, 
$\Delta log(\bar{P}/P) \approx 0.33$ is similar to the root-mean-square of the 
uncertainty of each of the constituents, estimated by us as 0.32.

Before continuing , some remarks about the validity of the derived background are 
necessary. Inspection of the literature indicates a range of predictions, most notably
that of the exponent of the energy spectrum beyond some tens of GeV. It is self-evident
that it should be 'large' because the Galactic lifetime of CR varies as, approximately,
$E^{-0.5}$ and the lifetime factor appears twice in the $\bar{P}$ intensity but only 
once in the $P$ intensity. The steep energy dependence of the boron to carbon ratio 
(i.e.secondary to primary) is a case in point. Ours is similar to that given 
elsewhere \cite{Kohri}  
\subsection{Derivation of the energy spectrum of antiprotons produced by the Single 
Source}
The AMS-02 collaboration presented actually the ratio of antiproton and proton 
intensities, in which the antiproton intensity is the sum of the background and the 
Single Source constituents: $\frac{I_{\bar{P}}^{bgrd} + I_{\bar{P}}^{SS}}{I_P}$.
Our calculations of the background have given the $\frac{I_{\bar{P}}^{bgrd}}{I_P}$
ratio. Subtracting the latter from the former ratios we obtain 
$\frac{I_{\bar{P}}^{SS}}{I_P}$ ratio. Multiplying it by the experimental proton 
intensity $I_P$ from \cite{AMS1} we obtain the energy spectrum of the antiprotons 
$I_{\bar{P}}^{SS}$ produced {\em by the Single Source.} The result is shown in Figure 2
 in comparison with the energy spectrum of protons from the same Single Source taken
from \cite{EW7}.
\begin{figure}[htb]
\begin{center}
\includegraphics[height=15cm,width=8cm,angle=-90]{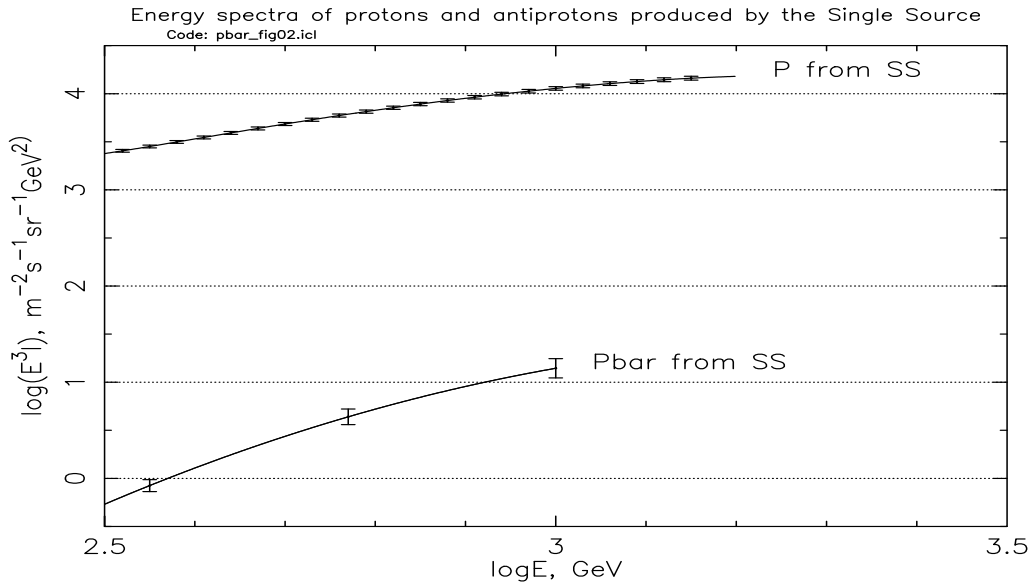}
\end{center}
\caption{Energy spectra of protons and antiprotons from the Single Source.}   
\label{fig:fig2}
\end{figure}   
 
Insofar as the $\bar{P}/P$ results are preliminary and there is a residual uncertainty 
in the estimate of the background the Single Source antiproton spectrum is imprecise. 
Nevertheless, the described calculations illustrate what should be eventually achieved.
It is seen that protons and antiprotons have similar spectral shapes. It is due to the
 constant $\bar{P}/P$ ratio and the dominant contribution of the Single Source to the
 total antiproton flux.
\subsection{On the way to the identification of the Single Source}
Using our model of the acceleration and propagation of CR \cite{EW4,EW5} together with 
calculations of the antiproton production by various primary CR nuclei \cite{Kach} we
calculated the possible contribution to $\bar{P}/P$ ratio of antiprotons generated by
three nearby SNR: Vela, Monogem and Geminga. All are comparatively close to the 
Earth ($\sim 300pc$), but have substantially different ages. Vela is young 
($\sim 10^4y$) and Geminga is much older ($\sim 3\cdot 10^5$y). Due to its old age, the
 SNR associated with Geminga has disappeared, but a very powerful pulsar has been 
preserved. We selected these SNR, because our previous analysis indicated that Vela can
 be the source 
responsible for the knee at PeV energies \cite{EW3} and Geminga is the candidate for 
the source of the positron excess \cite{EW4} at sub-TeV energies. The result of our 
estimates is shown in Figure 3. 
\begin{figure}[htb]
\begin{center}
\includegraphics[height=15cm,width=9cm,angle=-90]{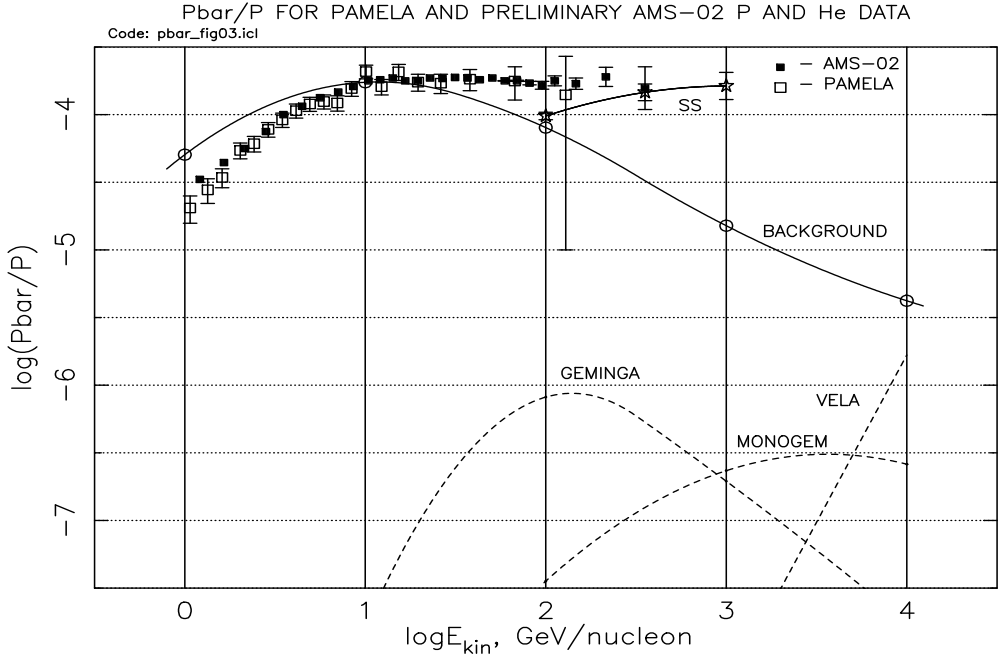}
\end{center}
\caption{Antiproton to proton ratio measured in PAMELA and AMS-02 experiments compared
with the possible contribution of three known sources: Vela, Monogem and Geminga SNRs. 
Contributions of the assumed Single Source and the background derived from the 
comparison of calculations and the experiment are denoted respectively.}   
\label{fig:fig3}
\end{figure}   
 
It is seen that the Vela SNR is so young that its accelerated protons and produced 
antiprotons of TeV energies have not reached the Earth and do not contribute to the  
$\bar{P}/P$ ratio. On the opposite side, Geminga is so old that all the sub-TeV 
particles
which can produce antiprotons have already gone. The shape and the absolute value of 
the Single Source contribution to the observed $\bar{P}/P$ ratio is such that this 
source should be much more powerful than Geminga and younger to explain the unusual
constancy of the ratio as a function of the energy. The Single Source should have
 characteristics 'intermediate' between these of Vela and Geminga. 

One of the possible candidates is the Monogem Ring SNR with its pulsar B0656+14. It is 
at about the same distance of $\sim$300pc from the Earth as Vela and Geminga, but its 
age is intermediate between them. Derived from the period and spindown rate of the 
associated pulsar, its age is $\sim 10^5$year. We have already discussed Monogem as
 a possible source responsible for the knee in the primary CR energy spectrum at 3-4 
PeV \cite{EW6}. This possibility arises, if during the SNR expansion, CR are confined 
inside the SN remnant for about $0.8\cdot 10^5$y and, in fact, emerged quite recently 
like CR from Vela. Recent more realistic models of SNR acceleration allow CR 
 to escape from the remnant from the very beginning after the SN explosion, especially 
at high energies. Therefore, we consider now that CR emerge and propagate during its 
entire age of $\sim 10^5$y for Monogem and $\sim 3\cdot 10^5$y for Geminga. In the 
process of propagation they collide with ISM atoms and produce antiprotons among other 
secondary particles. It is seen in Figure 3 that the Monogem gives flatter energy 
dependence of the $\bar{P}/P$ ratio in the 0.1-1 TeV energy range compared with Geminga
 which is closer to the results of the AMS-02 experiment.           

\section{Conclusions} 
Our calculations show that the dominant contribution to the antiproton production is 
given by proton-induced reactions. Primary helium nuclei contibute no more than 30\% of
 the total antiproton flux. The similarity of the produced antiproton and proton energy
 spectra derived for the Single Source can be explained if the Single Source is nearby 
and relatively recent to minimize the loss of high energy antiprotons due to the 
escape from the Galaxy.

Comparison with the measurements of the AMS-02 experiment shows that the
Single Source model and the potential nearby sources such as Monogem and Geminga can be
 responsible for the antiproton excess giving rise to the constant $\bar{P}/P$ ratio if
they are much more powerful than the standard SNR which converts $10^{50}$erg of its 
energy into CR. An alternative, and possible contributory effect, arises if the SNR is 
accompanied by dense molecular gas - a not uncommon fact. The yield of antiprotons 
could then be considerably higher. A similar situation exists for the explanation  of 
the results on the proton and helium spectra \cite{EW7}, where there was a shortage of 
one to two orders of magnitude in intensity; local gas has no relevance here, however. 
The Monogem SNR is better candidate than Geminga because it can 
give a flatter energy dependence of $\bar{P}/P$ ratio at sub-TeV energies. 
  
We conclude that with some reservations our Single Source model can give an adequate 
explanation of the preliminary $\bar{P}/P$ results from the AMS-02 experiment.

{\bf Acknowledgements} 

The authors are grateful to the Kohn Foundation for financial support.

\end{document}